\documentclass{elsart}
\usepackage{graphicx}
\usepackage{txfonts}

\begin{document}

\begin{frontmatter}

\title{Newtonian and Pseudo-Newtonian Hill Problem}

\thanks[mailsteklain]{steklain@ime.unicamp.br}
\thanks[mailletelier]{letelier@ime.unicamp.br}

\author{A. F. Steklain\thanksref{mailsteklain}}
\author{P. S. Letelier\thanksref{mailletelier}}

\address{Departamento                 de Matem\'atica Aplicada, Instituto de Matem\'atica, Estat\'{\i}stica e Computa\c{c}\~ao Cient\'{\i}fica,
Universidade Estadual de Campinas, 13083-970, Campinas, S\~ao Paulo, Brazil}

\begin{abstract}
A   pseudo-Newtonian   Hill problem based on the 
Paczy\'nski-Wiita pseudo-Newtonian potential that  reproduces 
 general relativistic effects is presented and compared with the usual
 Newtonian Hill problem.
Poincar\'e maps, Lyapunov exponents and fractal escape techniques are employed to study bounded and unbounded
 orbits. In particular we consider  the systems composed by
 Sun, Earth and Moon and composed by the Milky Way, the M2 
cluster and a star. We find that some pseudo-Newtonian systems - including the
M2 system - are more stable than their Newtonian  equivalent.
\end{abstract}


\end{frontmatter}

\section{Introduction}

In the 19th century G.W. Hill \cite{Hill} presented an approximation of the Moon-Earth-Sun system, where the movement of 
the Moon around the Earth was just perturbed by a distant   Sun. This
approximation became known  as the Hill Problem.
 Nowadays, this approximation is still applied in solar system models where
 bodies in nearly circular orbits are perturbed by other far away  massive 
bodies. Also, the Hill problem is very useful in the study of the stellar
dynamics, e.g.,  consider a star in a star cluster this last
     orbiting  around  a galaxy. The star, the cluster, and the galaxy can be
 considered as the Moon, the Earth, and the Sun, respectively. Despite the fact that the potentials of the cluster and 
the galaxy are far from being the
 potentials of a point masses  and their  orbits are
far from being circular  the Hill problem can be taken as a 
first approximation  and can easily accommodate necessary
 modifications, see for
instance, Heggie \cite{Heggie}.

The Hill Problem was first formulated in the realm of  Newtonian dynamics. 
However, there exists  extreme cases, like very  massive 
bodies, black holes, and  systems at great velocities, etc.  wherein the
Newtonian mechanics is no longer valid and relativistic corrections or a
fully general relativistic approach is needed. The Hill problem was proved to be non-integrable by 
Meletlidou et al. \cite{Meletlidou}, and is chaotic, as shown by Sim\'o and
Stuchi \cite{Stuchi}.

The aim  of this paper is first to study a
  pseudo-Newtonian Hill problem obtained replacing the Newtonian potential
   by the  Paczy\'nski-Wiita \cite{Paczynski} potential  that reproduces in a
  rather simple way  some aspects of the 
general relativistic dynamics. This is can be considered as 
a ``zeroth order'' approach to a general relativistic  Hill problem. A more
  rigorous approach should consider a schema of systematic approximations
  within  the realm of  Einstein theory of gravitation like the schema of
  post-Newtonian approximations. Due to the mathematical complexity of the
  post-Newtonian approximations in this case it is worth to begin with the
  simpler, but not rigorous, treatment of the Hill problem based 
on  the  pseudo-Newtonian potential above mentioned. The results obtained will
  be used as a starting point for a more complete treatment of the same problem
  in a future work.

 We shall compare the stability of orbits of the third body in the
 pseudo-Newtonian general relativistic simulation
with the equivalent classical Newtonian system. In particular we shall study
 the Sun-Earth-Moon system and  the three body problem associated with
the  Milky Way, the M2  cluster and a star.    

In the Section 2 we review  the Newtonian Hill problem, with
 its only integral of motion, the Jacobi integral.
  In Section 3 we present the pseudo-Newtonian Hill 
problem obtained obtained replacing the Newtonian potential
   by the  Paczy\'nski-Wiita   potential. In the Section 4 
 we  study the stability  of the third body (Moon)  orbits. 
 The techniques 
 used are:  Poincar\'e sections, fractal escapes and fractal dimensions, and 
 Lyapunov exponents. Finally, in  Section 5, we present our conclusions.      

\section{Hill's Equations of Motion}

\begin{figure}
\resizebox{\hsize}{!}{\includegraphics{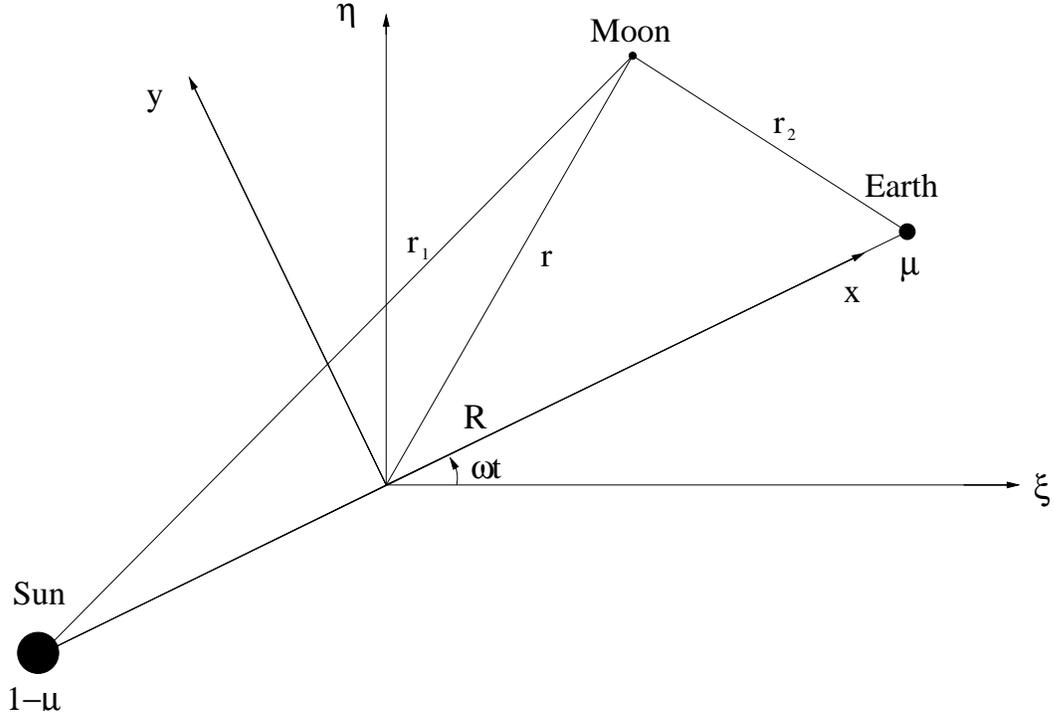}}
\caption{The planar, circular restricted three-body problem, on a
 inertial coordinate system $(\xi,\eta)$ and on a rotating coordinate
 system $(x,y)$. Note that on the rotating coordinate system the positions
 of the two massive bodies are fixed. }
\label{sistema}
\end{figure}

The Hill's lunar problem is a special case of the circular, planar restricted
 three-body problem, as described by Murray \cite{Murray} and Arnold
 \cite{Arnold}. In this problem there is a system of two massive bodies with
 masses $m_1$ and $m_2$, $m_2<<m_1$, in  circular orbits around their center
 of mass and a third massless body moving under influence of this system
 without perturbating it. We can choose units of mass such that
 $G(m_1+m_2)=1$. In this way, we can take
 the masses of the two massive bodies respectively as $1-\mu$ and $\mu$,
 where $\mu=Gm_2$ (note that in this units $m_2<m_1$ implies $\mu <1/2$) . If
 we consider the center of mass of the two massive bodies as the origin of 
the coordinate system, The massive bodies with masses $\mu$ and $1-\mu$ are 
situated respectively at the positions $(1-\mu)R$ and $\mu R$ of a rotating
 $x$-axis (see Fig.\ref{sistema}), where $R$ is the distance between
 the massive bodies. We have
\begin{equation}
R=\left( \frac{G(m_1+m_2)}{\omega ^2}\right)^{1/3}
\end{equation}
for this distance, where $\omega$ is the angular velocity of the rotating
frame. By  using the unit   of mass above defined and a
  unit of time such that $\omega =1$, we have $R=1$. In these units, the
  equations of the motion for the third body read
\begin{eqnarray}
\ddot{x} = 2\dot{y}+x -\frac{\partial V}{\partial x},
\\
\ddot{y} = -2\dot{x}+y -\frac{\partial V}{\partial y},
\end{eqnarray} 
where
\begin{equation}
V = -\frac{(1-\mu)}{r_1} - \frac{\mu}{r_2}
\end{equation}
with
\begin{eqnarray}
r_1=\sqrt{\left( x+\mu \right)^2 + y^2},
\\
r_2=\sqrt{\left[ x-(1-\mu)\right]^2+y^2}.
\end{eqnarray}
To obtain Hill's equations, we perform the transformation
\begin{eqnarray}
x \rightarrow (1-\mu) + \mu ^{1/3}x,
\\
y \rightarrow \mu ^{1/3}y.
\end{eqnarray}
This transformation corresponds to transfer the origin of the coordinate
 system to the body of mass $\mu$ and to consider the motion in a disk of
 radius  $\mu ^{1/3}$ centred  in this body. 
By considering $\mu$ small
 and taken  in expansion terms up to  $\mathcal{O}(\mu ^{1/3})$, we
 obtain the Hill's equations
\begin{eqnarray}
\ddot{x} = 2\dot{y}+ 3x -\frac{x}{r ^3},
\\
\ddot{y} = -2\dot{x} -\frac{y}{r ^3},
\end{eqnarray}
where $r = \sqrt{x^2 + y^2}$. These   equations can also be written
 as
\begin{eqnarray}
\ddot{x}-2\dot{y} = -\frac{\partial U}{\partial x},
\\
\ddot{y}+2\dot{x} = -\frac{\partial U}{\partial y},
\end{eqnarray}  
where $U$ is the Hill potential
\begin{equation}
U = -\frac{3}{2}x^2 - \frac{1}{r}.
\end{equation}
  The Hill equations can also 
be obtained from the Hamiltonian
\begin{equation}
H = \frac{1}{2}\left(p_x^2 + p_y^2\right) - \frac{1}{r} + \left(yp_x-xp_y\right) + \frac{1}{2}\left(y^2-2x^2\right),
\end{equation}
where $p_x=\dot{x}-y$ and $p_y=\dot{y}+x$ are the generalized momenta. 
$H$ is the Jacobi Integral, and can be written in the form
\begin{equation}
C_J = H = \frac{1}{2}\left( \dot{x}^2+\dot{y}^2 \right) - \frac{1}{r} - \frac{3}{2}x^2.
\end{equation}

Meletlidou et al. \cite{Meletlidou} proved that $C_J=H$ is the 
only motion integral of the Hill equations, therefore  the Hill  problem is 
non-integrable.

\section{A pseudo-Newtonian Hill problem}

 The restricted three-body problem  with  general relativistic corrections
(post-Newtonian corrections) is unknown, and as in the simpler case of the 
 two body problem it should  lead to equations that are not simple  for a first
 approach. By this reason we shall simulate general relativistic effects
via  the  pseudo-Newtonian potential introduced by  Paczy\'nski-Wiita 
\cite{Paczynski} to the study of accretion disks in 
black holes.  Also this potential has been used to study black holes with
halos by  Gu\'eron and Letelier \cite{Gueron}).  
Other pseudo-Newtonian models can be found in
literature, e.g., the one studied by Semer\'{a}k and Karas \cite{sk} to
describe rotating black holes.
 The potential introduced by  Paczy\'nski-Wiita  
\cite{Paczynski}  changes 
the usual $GM/r$  Newtonian potential by 
$GM/(r-r_S)$, where $r_S$ is the Scharzschild radius, as in the General 
Relativity, $r_S=2GM/c^2$. This potential exactly reproduces the
 marginally bound circular
orbit, $r_{mb} =2r_S$, and the last stable circular 
orbit, $r_{ms}=3r_S$, and yields efficiency factors
 in close agreement with the Schwarzschild solution. In particular,
 in  the  system under  consideration,  the third body moves 
in  the vicinity of
 the second body with mass $\mu$.  The first body  with mass $1-\mu$
is far away from the second and third  bodies. Then we change the potential of
the second body by  $\mu/(r-r_{S2})$, and we kept the potential of the
 first  body, $(1-\mu)/(r-r_{S1}) \approx (1-\mu) /r$,  where
 $r_{S1}$ and $r_{S2}$ are the Scharzschild radius of the first 
                                        and second body, respectively.
 The modified Hill equations are
\begin{eqnarray}
\ddot{x}-2\dot{y} = 3x -\frac{x}{r(r-r_S^*)^2}= 
-\frac{\partial U_P}{\partial x},
\\
\ddot{y}+2\dot{x} =-\frac{y}{r(r-r_S^*)^2}= -\frac{\partial U_P}{\partial y},
\end{eqnarray}  
where $U_P$ is the modified Hill potential
\begin{equation}
U_P = -\frac{3}{2}x^2 - \frac{1}{r-r_S^*},
\end{equation}
and $r_S^*=\mu^{-1/3}r_{S2}$, in units such that the separation between the
 two massive bodies is  $R=1$, like in the Newtonian case. The modified Jacobi
 constant, in this case, is given by
\begin{equation}
C_{JP} = \frac{1}{2}\left( \dot{x}^2+\dot{y}^2 \right) - \frac{1}{r-r_S^*} - \frac{3}{2}x^2.
\end{equation}
From the condition that  the horizons of 
 the massive bodies do  not  intersect we have an  upper limit for  $r_S^*$. 
 This implies $r_{S1} + r_{S2}< R$. From the condition 
 $\mu <1/2$  
 we find (after some algebra) that  $r_S^*<2^{-2/3}\approx 0.63$.  

\section{Stability of orbits}

We compare the Newtonian system ($r_S^*=0$) and pseudo-Newtonian systems
 with parameters $r_S^*$ varying from $10^{-12}$ to $10^{-3}$. The 
value $r_S^*=4 \times 10^{-12}$ corresponds to the Sun-Earth-Moon system, and
  $r_S^*=5 \times 10^{-10}$  to a system involving the Milky Way, the M2 cluster and
 a typical star. The others represent systems with a  very massive
 second body.    

\subsection*{ a) Poincar{\'e} sections}

\begin{figure} 
\includegraphics[totalheight=0.2\textheight,viewport=-20 0 690 558,clip]{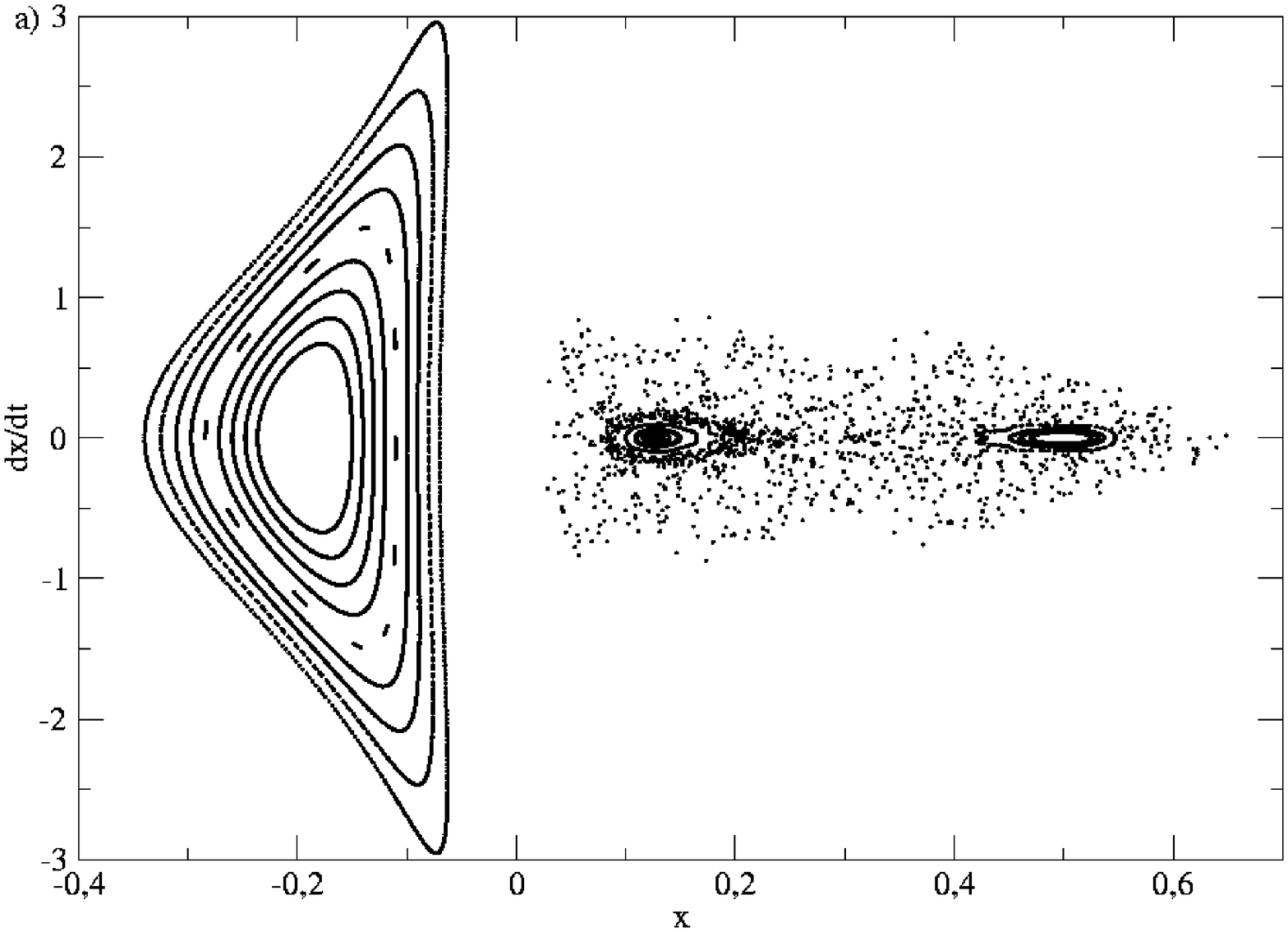}
\hfill
\includegraphics[totalheight=0.2\textheight,viewport=-20 0 690 558,clip]{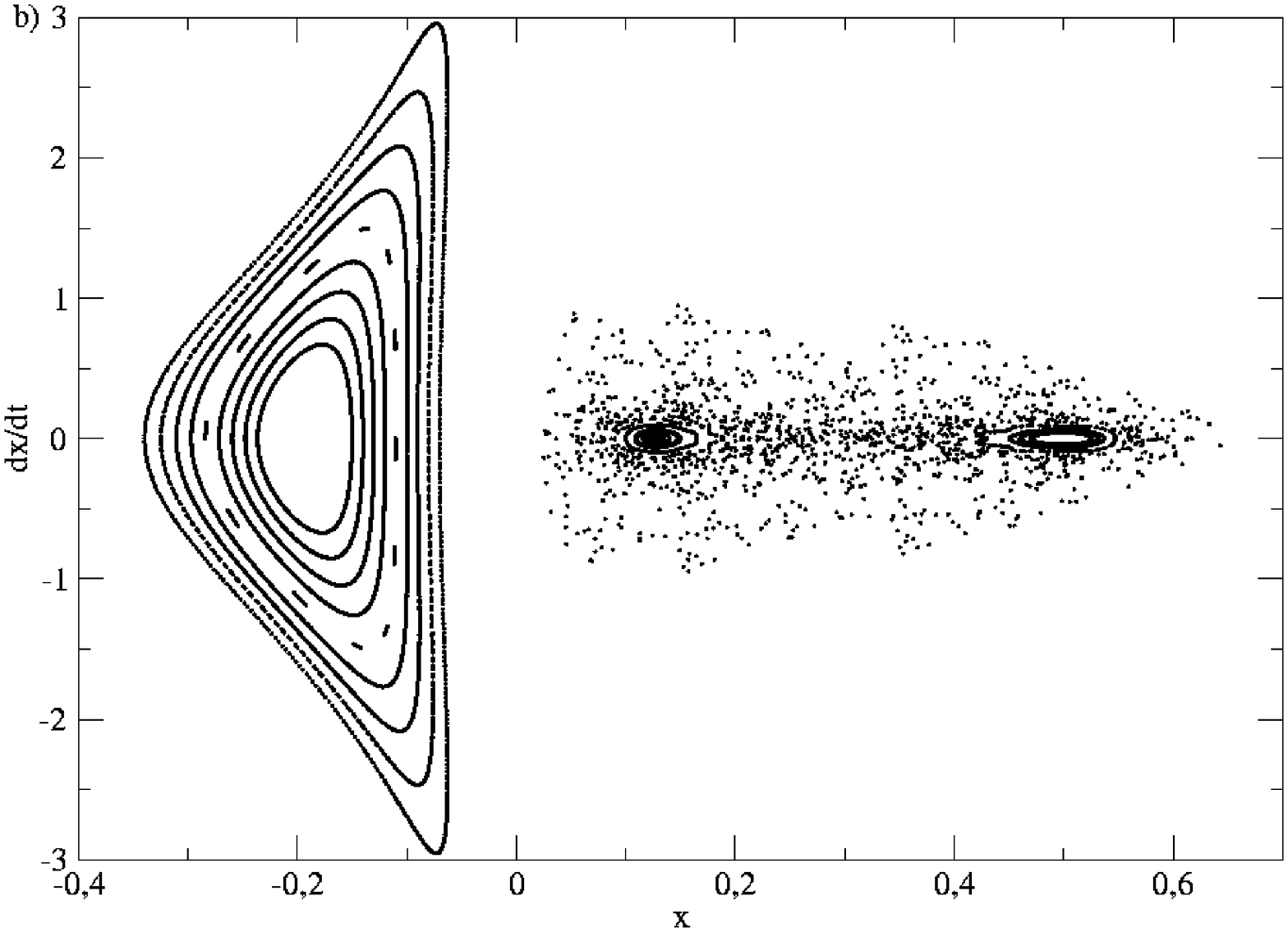}
\hfill
\includegraphics[totalheight=0.2\textheight,viewport=-20 0 690 558,clip]{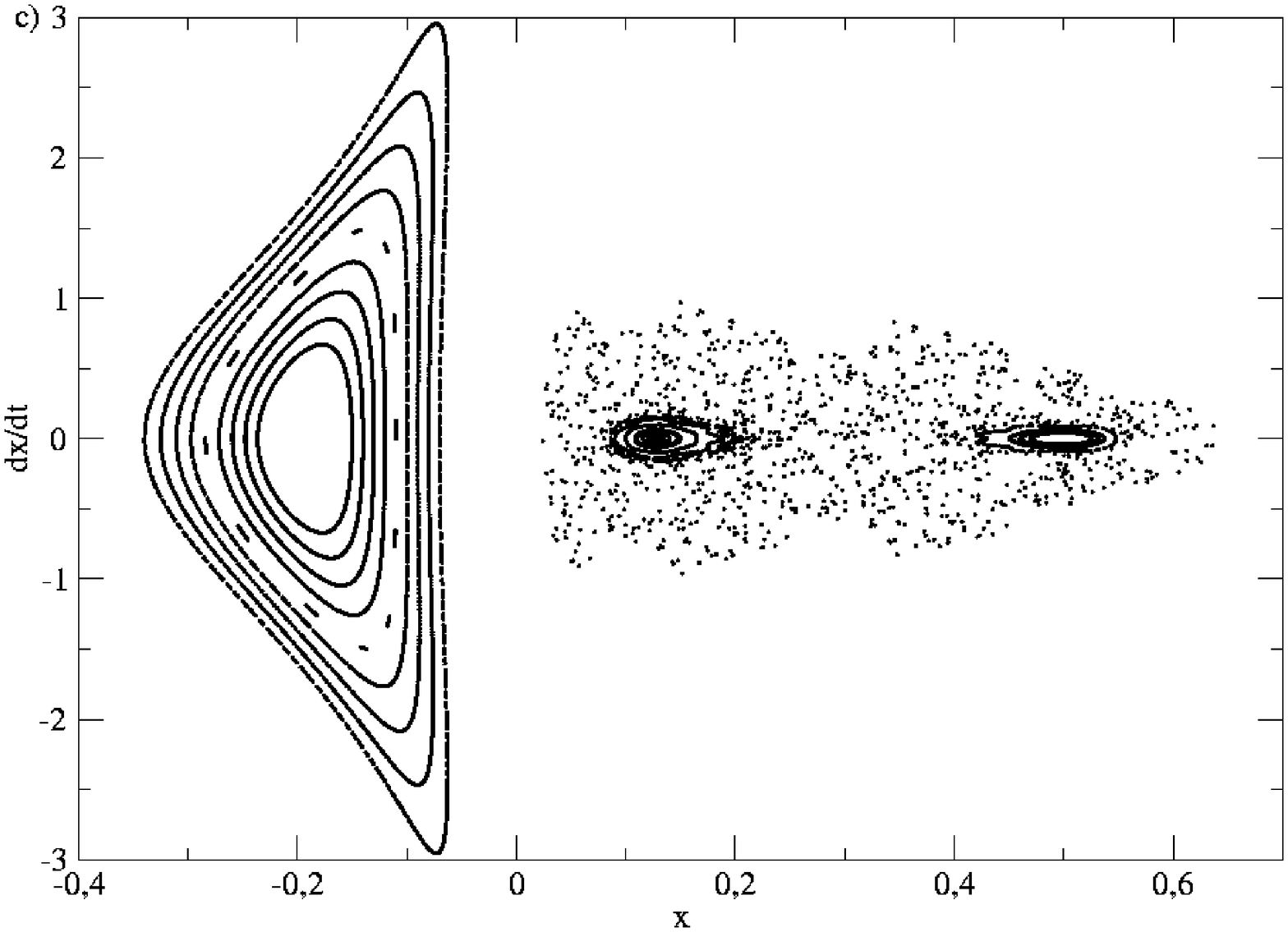}
\hfill
\includegraphics[totalheight=0.2\textheight,viewport=-20 0 690 558,clip]{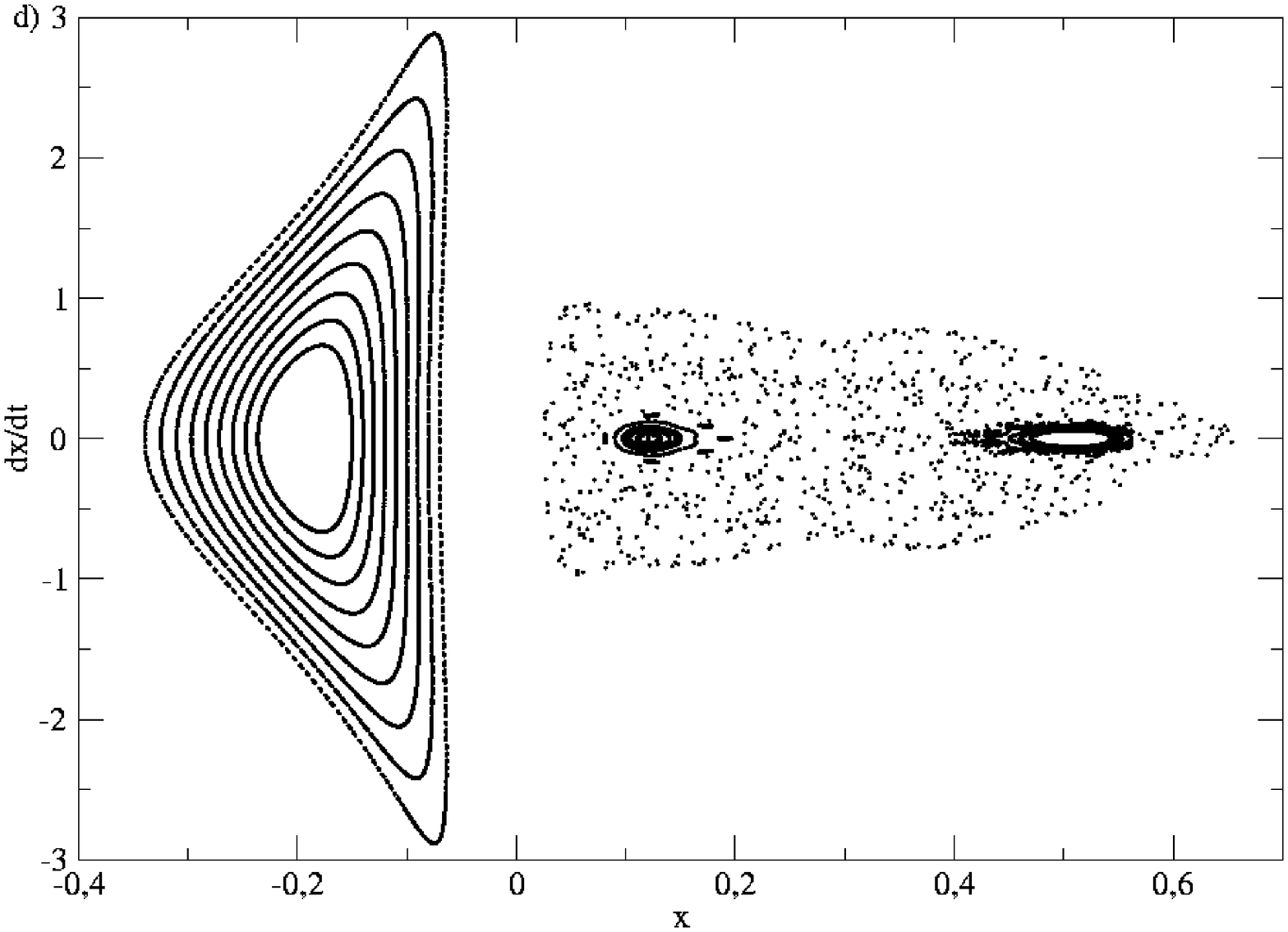}
\hfill
\caption{Poincare sections for a) the Newtonian system ($r_S^*=0$) , b) pseudo-Newtonian system with $r_S^*=5 \times 10^{-10}$, c) $r_S^*=5 \times 10^{-6}$ and d) $r_S^*=5 \times 10^{-4}$. The figures a) and c) are very similar, and in d) it can be seen the formation of island chains around the stable region in the middle.}
\label{poincarec}
\end{figure}

The orbits in the Newtonian as well in the pseudo-Newtonian Hill problem
are the solutions of a four dimensional dynamical system with variables
 $(x,y,\dot x,\dot y )$. Since we have an integral of motion, $C_J$, the
motion is reduced to a  three dimensional system, we can take 
 $(x, \dot x , y)$ as independent variables.
We shall study surface of section (Poincar\'e sections)
 evaluating the 
orbits for different values of the Jacobi constant and registering
 the crossings of the hypersurface $y=0$ with $\dot y > 0$. The results
 for $C_J= -2.17$ are shown on the Fig. \ref{poincarec}. We see that 
  some  Kolmogorov-Arnold-Moser (KAM)  tori are  destroyed, indicating the
 transition of the system from regular to chaotic behaviour. Note 
that the picture for $r_S^*=0$ and for $r_S^*=5 \times 10^{-6}$ are 
similar.  Another remarkable feature of this maps is the 
presence, for $r_S^*=5 \times 10^{-4}$ case, of island chains. These structures
usually  
give rise to more destroyed tori indicating  less stability of the system.

\subsection*{b) Lyapunov exponents}

We shall study  the Lyapunov exponents for the systems above described  to
better analyze  the orbits stability. We shall use the Lyapunov
characteristic number ($\lambda$) that is defined as the double limit 
 \begin{equation}\label{lcn}
\lambda=\lim_
{\scriptsize
\begin{array}{l} \delta_0\rightarrow 0 \\
 t\rightarrow \infty 
 \end{array}
}
 \left[\log(\delta /\delta _{0}) \over t\right] , 
\end{equation}
 where $\delta _{0}$ and  $\delta $\ are the deviation of
two nearby orbits at times $0$ \ and $t$\, respectively (see Alligood et al. \cite{Yorke}). We get
the largest $\lambda$ using the technique suggested by Benettin et al.
 \cite{Galgani}, in particular we use an  algorithm due to
  Wolf et al. \cite{Wolf}.

The Lyapunov exponents are not absolute, but dependent 
on the choice of the time scale. We recall that  we have fixed 
 the time scale by the requirement that $\omega=1$.  This  defines a time unit
that is natural to each particular system. In these time  units,
 for two different systems with equal Lyapunov exponents measured in inverse
 seconds  and  different angular frequencies 
  $\omega_1$ and $\omega_2$ such that $\omega_1 <\omega_2$,
we will have that  the system with
 larger frequency  will have small   Lyapunov exponent in the natural units of
 time, and vice versa.
 In other words, in this case,  the separation of nearby orbits of the third
 body  in each
 revolution of the  first body will be larger for the system with smaller 
$\omega$.
So we shall consider that, even though the two systems have equal  Lyapunov
exponents measured in inverse  seconds the system with smaller angular
frequency is more unstable that the one with larger $\omega$.

\begin{figure}
\resizebox{\hsize}{10cm}{\includegraphics[totalheight=0.2\textheight,viewport=-20 0 690 558,clip]{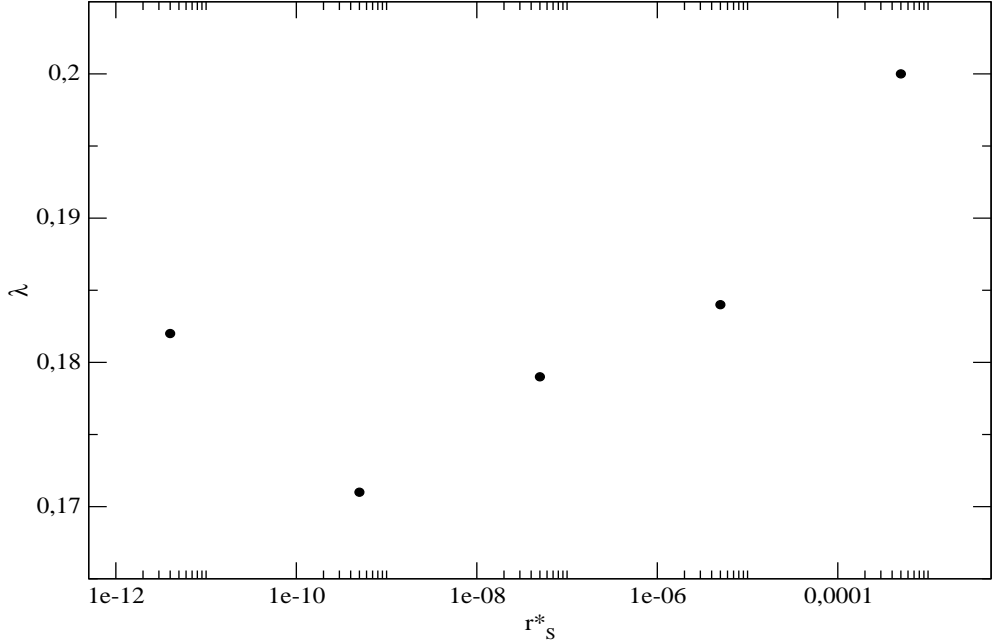}}
\caption{Lyapunov exponents for different values of $r_S^*$. Note that there is a minimum around  $r_S^*=5 \times 10^{-10}$.} 
\label{lyapunov}
\end{figure}

For $r_S^*=0$  we get  $\lambda=0.185 \pm 0.003$. The Lyapunov
 exponents  for 
 $r_S^*=    4\times  10^{-12},  5 \times  10^{-10} ,5\times 10^{-8}, 
  5\times  10^{-6}, 5\times 10^{-4}  $ are shown in  the Fig. \ref{lyapunov}. 
  We  have a local minimum around $r_S^*=5\times 10^{-10}$.

\subsection*{c) Fractal escape and fractal dimension}

\begin{figure}
\centering
\resizebox{\hsize}{10cm}{\includegraphics[totalheight=0.2\textheight,viewport=-20 0 400 420,clip]{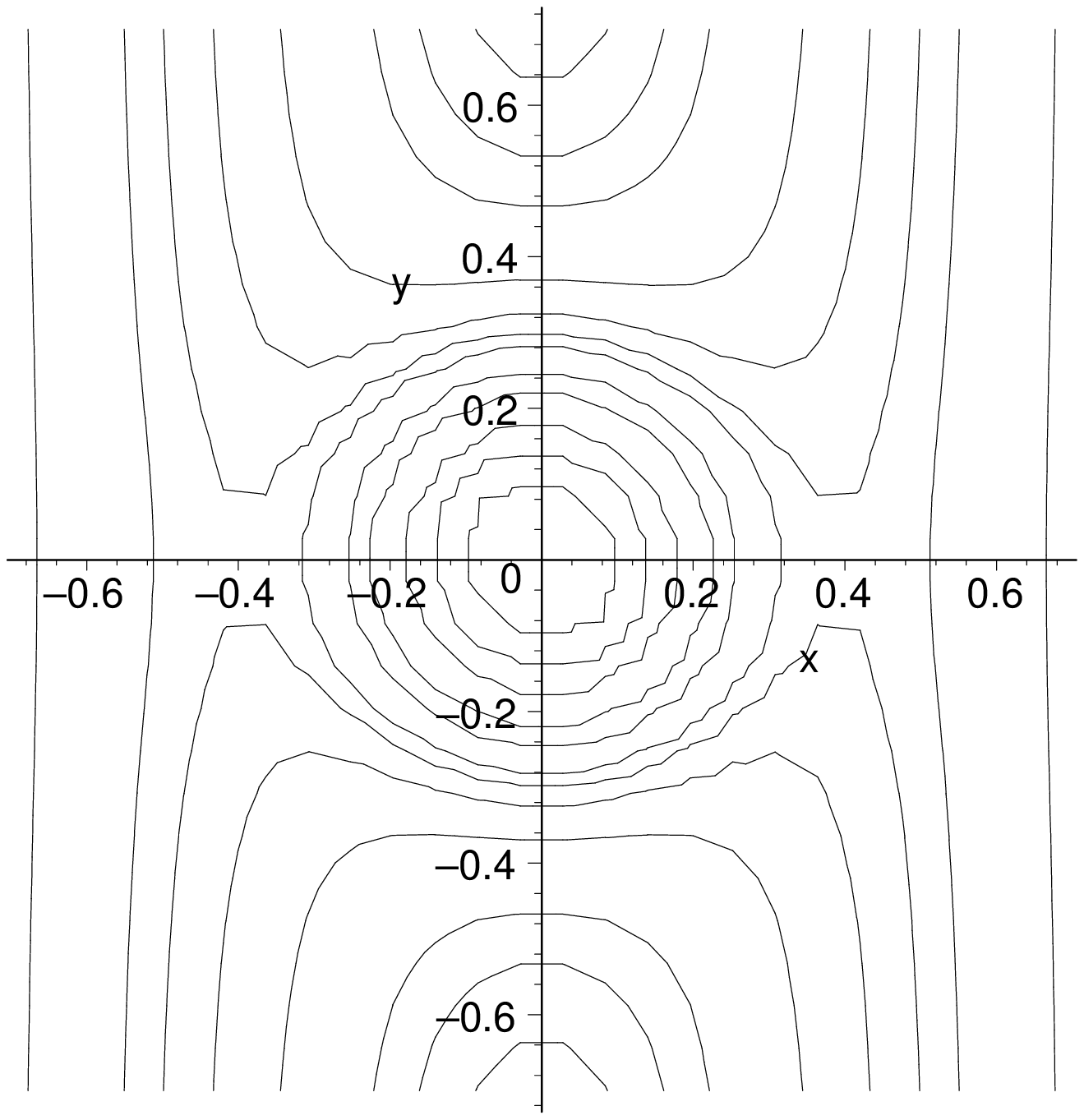}}
\caption{Contour plot of the potential $U$ for  the Newtonian system. For
 ``energies''
 above $-2.16$ there are two escape
 routes, $x \rightarrow +\infty$ and $x \rightarrow -\infty$.} 
\label{potential}
\end{figure}

The Poincar\'e sections were obtained for values of the Jacobi 
constant such that the
 systems are  bounded. For values larger than 
$C_{Jbounded}= -2.16$ the systems  are unbounded and the third body escapes. In Fig. \ref{potential}
we compute contour plot   for the 
potential $U$  of the Newtonian system. The  pseudo-Newtonian
system is quite similar and will not be presented her. We see
 two escape routes, $x \rightarrow +\infty$ and $x \rightarrow -\infty$.

\begin{figure}
\resizebox{\hsize}{!}{\includegraphics[totalheight=0.2\textheight,viewport=-20 0 690 558,clip]{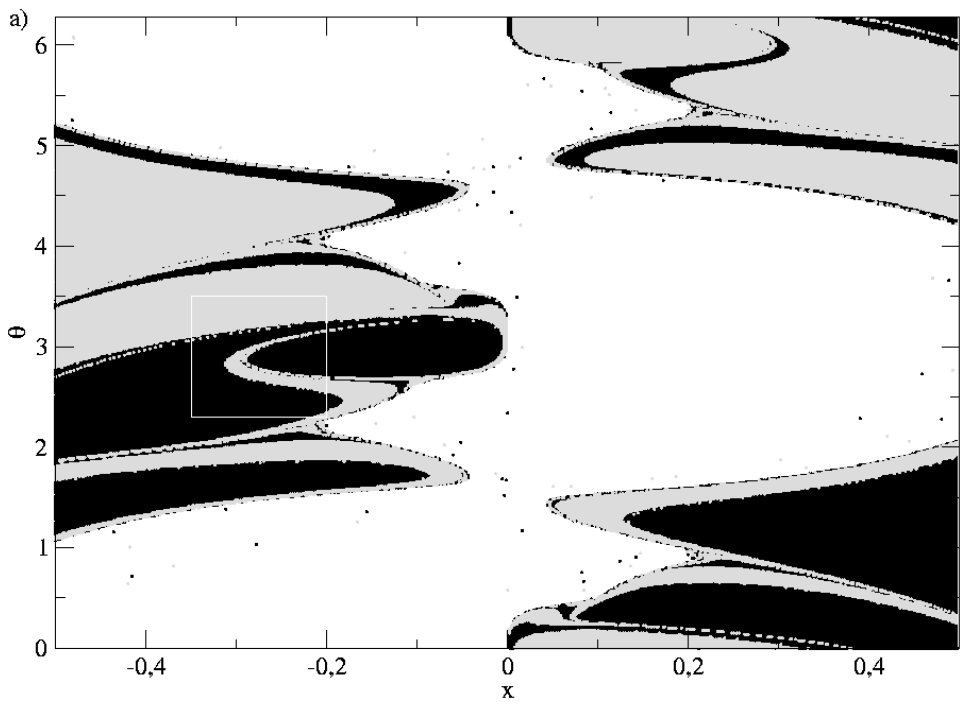}}
\hfill
\resizebox{\hsize}{!}{\includegraphics[totalheight=0.2\textheight,viewport=-20 0 690 558,clip]{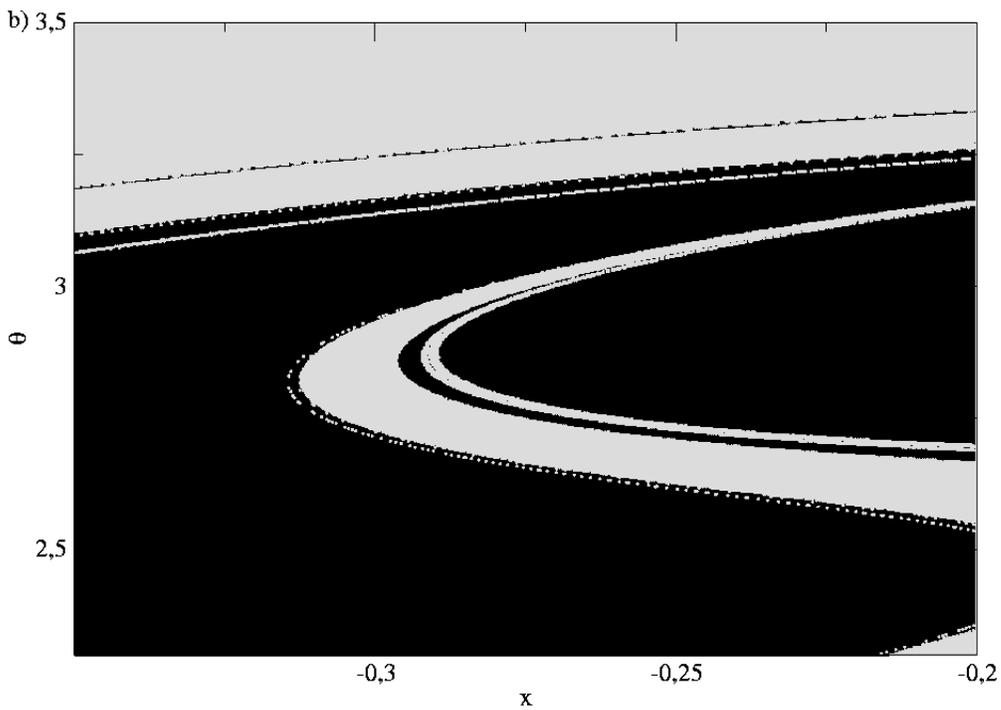}}
\caption{(a) Escape basins for the Newtonian system.
Grey (black)  means initial
 conditions for particles that  escape to 
$x \rightarrow +\infty$ ( $x \rightarrow -\infty$) and
white are the initial conditions for particles that do not
 escape during the integration time.  (b) A zoom of the portion of the precedent figure enclosed by
the square.} 
\label{basinc}
\end{figure}

 For open systems that have more than one route
 to  escape we can apply
 the fractal escape technique used by  Moura and Letelier \cite{Moura} in
 the study of the classical H\'enon-Heiles problem.  In
 this method the basins of the escape routes are obtained for a set
of initial conditions.  For chaotic systems, we  have the existence of
 fractal basin boundaries (FBB)  indicating a great instability of the orbits.
 In our case we chose a subset of the
 accessible  phase space at a fixed Jacobi
constant, defined by a segment $|x|\leq a$, $y=b$ and
  $0 \leq \theta \leq   2\pi$, where
 $a$ and $b$ are constants to be chosen appropriately
and $\theta$ is the angle that defines the direction of the  velocity with
 respect the $x$-axis. Then the trajectories are  integrated numerically,
  we have  three different cases: (1) the body escapes
 to $x \rightarrow +\infty$,
(2) it  escapes to $x \rightarrow -\infty$, and (3) the particle does  not scape
 during the integration time. We take  the  integration time long enough 
to be sure that our  results do not depend on this time.
Fig. \ref{basinc}(a)   shows the basins obtained for the Newtonian case. The
 pseudo-Newtonian basins are very similar and will not be shown here. 
In  the figures, grey means initial
 conditions for particles that  escape to 
$x \rightarrow +\infty$, black means that the particles escape to $x \rightarrow -\infty$ and 
white are the initial conditions for particles that do
not  escape. 
 A zoom of a portion of  \ref{basinc}(a) is shown in 
 Fig. \ref{basinc}(b), we see    self-similarity of the basin boundaries,
   a fractal characteristic.
    
To show the diference between  systems with different values of the parameter
 $r_S^*$, we calculate  the 
 dimension of the basin boundaries obtained for the different  systems. 
The dimension used is the  box-counting dimension that can be easily obtained, 
  see for instance,  Ott \cite{Ott} and  Grebogi et al. \cite{Grebogi}.
If we displace a 
determined point of a basin to another on a distance $\epsilon$, the
 probability 
that this new initial condition does not  belong to the same basin of the
 old one is, for small $\epsilon$, $P(\epsilon)\propto \epsilon^{D-d}$, where
 $D$ 
is the dimension of the set ($2$ in our case) and $d$ is the box counting
 dimension, also
 called exterior or fractal dimension when  not integer. 
   In order to calculate this fractal dimension, 
 for several values of $\epsilon$, we displace the $x$ coordinate of all
 the points from one of the basins ($x \rightarrow +\infty$), and count 
the number of points that does not  belong to the same basin. Then
we compute the   the fraction of
 numbers that does not  belong to the same basin, $P(\epsilon)$.  
 We plot $\ln P(\epsilon)$  in function of
 $\ln\epsilon$.  The inclination of the straight line gives us $D-d$.
 The obtained values for the fractal dimension, $d$, are shown in 
the Table~\ref{tab}. Some of the values  obtained are
 indistinguishable one from each other in  the precision used. Due to his high
 instability, we were not able to calculate the fractal dimension 
of the $r_S^*=4 \times 10^{-4}$ system. This value is probably near $2$, the higher value.

\begin{table}
\caption{\label{tab}Fractal dimensions for Newtonian and
 pseudo-Newtonian systems}
\centering
\begin{tabular}{c c}
\hline\hline
$r_S^*$ & Fractal Dimension \\
\hline
$0$ (Newtonian system) & $1.452 \pm 0.003$ \\
$4\times 10^{-12}$ & $1.451 \pm 0.003$ \\
$5\times  10^{-10}$ & $1.452 \pm 0.003$ \\
$5\times 10^{-8}$ &  $1.459 \pm 0.003$ \\
$5\times 10^{-6}$ &  $1.466 \pm 0.003$ \\
\hline
\end{tabular}
\end{table}

\section{Conclusions}
 We  can see  directly from the Lyapunov exponents and confirmed by 
the Poincar\'e sections that the pseudo-Newtonian systems  (a relativistic 
simulation)  are  not always more unstable than
 their equivalent  Newtonian systems.

 It could be guessed a priori that the
 pseudo-Newtonian systems should be more unstable  due to the fact that the 
Paczynski-Wiita potential introduces a saddle point in  the  dynamical 
system. However this  
happens only for $r_S^*>(9/4)^{1/3} \approx 1.31$.
As we mention before, due to the approximation used for the Hill problem,
 we have
$r_S^*< 0.63$.  Thus in the cases under
 consideration   the number of saddle points is the same for both Newtonian 
and 
 pseudo-Newtonian systems. Another feature is the presence of a local 
minimum for the Lyapunov  exponents, indicating that there exists 
 a more stable
 configuration for pseudo-Newtonian systems that is even more stable than the 
equivalent Newtonian configuration. This minimum corresponds to
 a known physical system, the  Milky Way-M2-star system. This is an 
evidence that relativistic effects can made a system more stable. 
This  is  related to the fact   that the addition of  
 extra spherical terms in the
 Newtonian potential can be used to  damp the influence of the non-spherical
 perturbation, making the motion more regular, see, for instance, Ivanov et al. \cite{Saha}. 
This needs to be confirmed by a fully General Relativistic treatment of the
Hill problem. We hope to comeback to this subject soon.

\section{Acknowledgements}
We want to thank CNPq and FAPESP for financial suport.

\end{document}